\documentclass[reprint, showpacs, amsmath, amssymb, superscriptaddress, prb, aps]{revtex4-1}

\usepackage[latin9]{inputenc}
\usepackage[T1]{fontenc}
\usepackage{color}
\usepackage{amsmath}
\usepackage{amsfonts}
\usepackage{amssymb}
\usepackage{sansmath}
\usepackage{units}
\usepackage{nicefrac}
\usepackage{graphicx}

\usepackage{setspace}	
\usepackage{textcomp}

\usepackage{bm}

\begin{document}

%\preprint{APS/123-QED}
%
\title{Large Negative Electronic Compressibility of LaAlO$_3$-SrTiO$_3$ Interfaces with Ultrathin LaAlO$_3$ Layers}% Force line breaks with \\
\author{V. Tinkl}
\affiliation{Center for Electronic Correlations and Magnetism, University of Augsburg, Universit\"{a}tsstra{\ss}e 1, 86135 Augsburg, Germany}
\author{M. Breitschaft}
\affiliation{Center for Electronic Correlations and Magnetism, University of Augsburg, Universit\"{a}tsstra{\ss}e 1, 86135 Augsburg, Germany}
\author{C. Richter}
\affiliation{Center for Electronic Correlations and Magnetism, University of Augsburg, Universit\"{a}tsstra{\ss}e 1, 86135 Augsburg, Germany}
\affiliation{Max Planck Institute for Solid State Research, Heisenbergstra{\ss}e 1, 70569 Stuttgart, Germany}
\author{J. Mannhart}
\email[]{J.Mannhart@fkf.mpg.de}
\affiliation{Max Planck Institute for Solid State Research, Heisenbergstra{\ss}e 1, 70569 Stuttgart, Germany}

\date{\today}

\begin{abstract}
A two-dimensional electron liquid is formed at the \textit{n}-type interface between SrTiO$_3$ and LaAlO$_3$. Here we report on Kelvin probe microscopy measurements of the electronic compressibility of this electron system. The electronic compressibility is found to be negative for carrier densities of $\approx10^{13}$\,/cm$^2$. At even smaller densities, a metal--to--insulator transition occurs. These local measurements corroborate earlier measurements of the electronic compressibility of LaAlO$_3$-SrTiO$_3$ interfaces obtained by measuring the capacitance of macroscopic metal-LaAlO$_3$-SrTiO$_3$ capacitors.
\end{abstract}

 

\maketitle

It has been established that a two-dimensional (2D) sheet of mobile electrons is generated at the interface between the TiO$_2$-terminated (001) surface of SrTiO$_3$ and LaAlO$_3$.\cite{ohtomo,basletic,caviglia} This electron system has remarkable properties that differ significantly from the properties of 2D electron gases embedded in semiconductor heterostructures. For example, the characteristic carrier density $n$ at LaAlO$_3$-SrTiO$_3$ interfaces equals several $10^{13}$\,/cm$^2$, which is well above the typical densities of $10^{11}$ - $10^{12}$\,/cm$^2$ found in semiconductor heterostructures. The charge carriers at the interface originate from an electronic reconstruction\cite{nakagawa,salluzzo} and occupy Ti~3$d~t_\mathrm{2g}$ states at the interface TiO$_2$ layer.\cite{breitschaft} For the effective mass of the electrons values in the range of 1 to 3.2 bare electron masses were reported.\cite{caviglia,pavlenko,shalom,dubroka} In the samples investigated the electron mobility is of order 1\,000\,cm$^2$/(Vs) at 4.2\,K.\cite{thiel_field} Moreover the system shows coexistent superconductivity and magnetism if cooled to low temperatures.\cite{dikin,luli_coex,bert} Furthermore, by performing capacitance measurements on SrTiO$_3$-LaAlO$_3$-Au and SrTiO$_3$-LaAlO$_3$-YBa$_2$Cu$_3$O$_{7-x}$ capacitors, a state with negative electronic compressibility $\kappa=(n^2\mathrm{d}\mu/\mathrm{d}n)^{-1}$ was identified, where $\mu$ is the electrochemical potential.\cite{luli_kap} In a dilute electron gas, a negative electronic compressibility results from the dominance of exchange and correlations terms, which apparently explain the negative compressibility in some semiconductor heterostructures.\cite{eisenstein_prb} The origin of the observed negative compressibility of the less diluted electron liquid at LaAlO$_3$-SrTiO$_3$ interfaces has not been identified completely. The negative electronic compressibility of the LaAlO$_3$-SrTiO$_3$ interface electron system was found to exceed the negative compressibility of 2D electron gases in Si heterostructures by a factor of at least ten.\cite{kravchenko} The negative electronic compressibility is also much larger than the negative compressibility recently reported in carbon nanotubes and GaAs structures.\cite{ilani,eisenstein,allison} Although all studies of the negative electronic compressibility at LaAlO$_3$-SrTiO$_3$ interfaces were done with samples in which the LaAlO$_3$ layers were 10 or 12\,unit cells thick to prevent tunneling currents, which are unfavorable in capacitance measurements, the negative electronic compressibility has been predicted to occur also in samples with LaAlO$_3$ films as thin as four monolayers, the thinnest films to generate a conducting LaAlO$_3$-SrTiO$_3$ interface.\cite{thiel_field} Because tunneling and leakage currents through the LaAlO$_3$ layer undermine the accuracy of measurements of the electronic compressibility using planar capacitors, we have explored such samples by local measurements of the electronic compressibility. For this we used Kelvin probe microscopy, which has previously been employed on LaAlO$_3$-SrTiO$_3$ heterostructures to map the distribution of the surface potential.\cite{kala} With Kelvin probe microscopy it is possible to measure the electronic compressibility of samples grown without a top gate, which would be detrimental to the studies of LaAlO$_3$-SrTiO$_3$ heterostructures with 4\,unit-cell-thick LaAlO$_3$ layers.\cite{luli_kap,arras,foerg}

To measure the electronic compressibility of LaAlO$_3$-SrTiO$_3$ interfaces we fabricated LaAlO$_3$-SrTiO$_3$ heterostructures comprising 4\,unit-cell-thick (1.6\,nm) epitaxial LaAlO$_3$ films. The LaAlO$_3$ films were grown by pulsed laser deposition on the TiO$_2$-terminated \cite{kawasaki,koster} (001) surface of SrTiO$_3$ crystals. The deposition was performed at a substrate temperature of 780\,°C, an oxygen background pressure of $\approx1\times10^{-4}$\,mbar, and was monitored by reflection high-energy electron diffraction. After cooling the samples in 400\,mbar O$_2$, an aluminum shadow mask with a rectangular hole was attached to the samples' surfaces. Using this mask, Ar ion etching was employed to etch holes into the SrTiO$_3$ substrates, which were then filled with electron-beam evaporated Ti to contact the interface electron liquid. After transporting the samples to the preparation chamber of the scanning probe microscope (SPM) they were heated radiatively for > 40 minutes to $\approx170$\,°C to clean their surfaces. The SPM, which operates in ultrahigh vacuum at 4.7\,K, utilizes a cantilever based on a quartz tuning fork (qPlus-sensor) \cite{giessibl} with a spring constant of $\approx1\,800$\,N/m. An iridium spall treated \textit{in situ} by field emission was used as a tip. The experimental setup is sketched in Fig. 1.

\begin{figure}
  \includegraphics[width=\columnwidth]{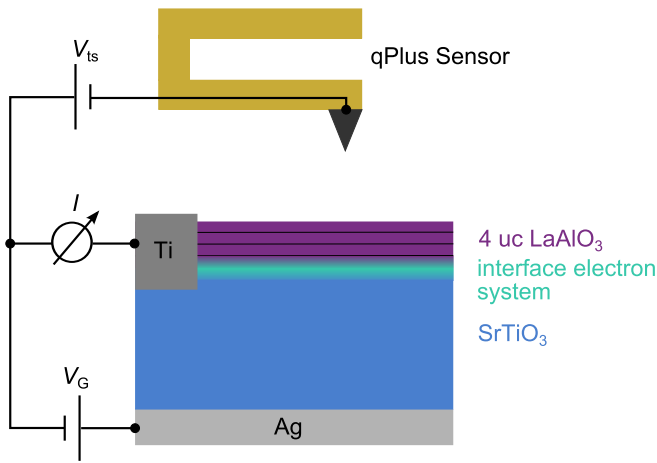}
  \caption{Illustration of the experimental configuration. The force sensor consists of a quartz tuning fork.\cite{giessibl} The force between tip and sample is monitored as a function of $V_{\mathrm{ts}}$, the voltage applied between tip and sample. Through the back-gate voltage $V_{\mathrm{G}}$ between the interface and the silver paint at the back of the SrTiO$_3$ substrate the carrier density at the interface can be tuned. A negative $V_{\mathrm{G}}$ causes the \textit{n}-type interface electron system to be depleted. The current $I$ flowing into the interface and the force between tip and sample are recorded simultaneously.}
  \label{fig:aufbau}
\end{figure}
The standard step--and--terrace structure of the LaAlO$_3$-SrTiO$_3$ heterostructures resulting from a slight vicinal cut of the SrTiO$_3$ substrates ($\approx0.15$\,°) is readily imaged by scanning force microscopy (Fig. 2). Whereas on more standard samples excellent resolution was easily achieved with the SPM,\cite{hembacher} it was not possible to obtain atomic resolution on the LaAlO$_3$-SrTiO$_3$ heterostructures.

\begin{figure}
  \includegraphics[width=\columnwidth]{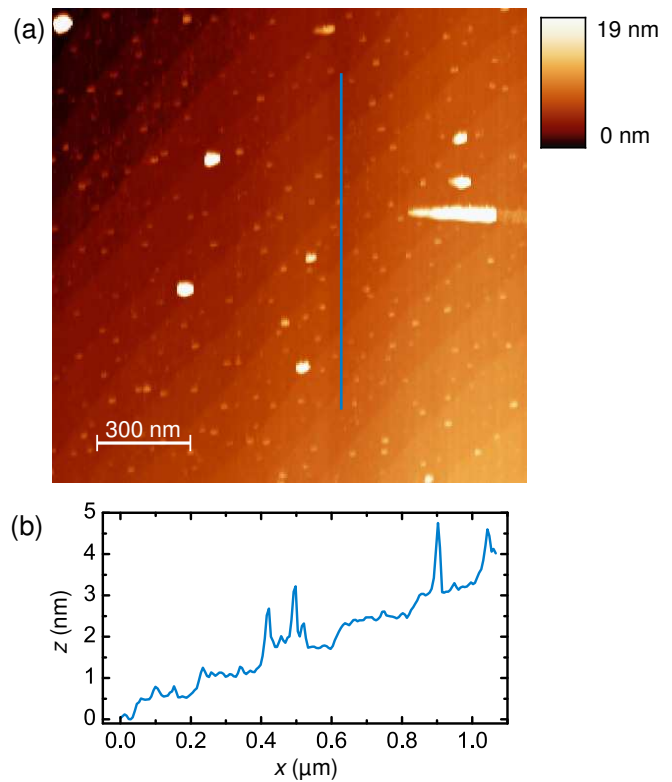}
  \caption{(a) Topographic frequency-modulation scanning force microscopy image of the LaAlO$_3$ film acquired at 4.7\,K. The image was recorded with a scanning speed of 75\,nm/s and a frequency shift of $\Delta f = -1.8$\,Hz. The free resonance frequency of the cantilever was $f_{0} = 25\,926.6$\,Hz, the quality factor was $Q = 20\,320$ and the oscillation amplitude was set to $A = 3.4$\,\AA. The white dots in the scan are caused by adsorbates on the sample surface.(b) Profile taken along the blue line plotted in (a).}
  \label{fig:scan}
\end{figure}
To assess the electronic compressibility of the interface we measured the force between the tip and the sample as function of tip--sample distance and tip--sample voltage. The electrostatic force $F_{\mathrm{es}}$ exerted on a tip by an electric field between the tip and a conducting sample is given by
\begin{equation}
F_{\mathrm{es}}(V) = -\frac{1}{2} \frac{\mathrm{d}C_{\mathrm{ts}}}{\mathrm{d}z} V^2,
\end{equation}
where $C_{\mathrm{ts}}$ is the capacitance of the tip--sample arrangement, $z$ the tip--sample distance, and $V$ the difference of the electric potentials of tip and sample. Tip and sample are uncharged for $V=0$.

If the work functions of tip and sample, $\phi_{\mathrm{t}}$ and $\phi_{\mathrm{s}}$, differ, electrons move between the tip and the sample to equilibrate the electrochemical potentials of tip and sample, $\mu_{\mathrm{t}}$ and $\mu_{\mathrm{s}}$. As a result, tip and sample become charged and a difference $V_{\Delta\phi} = \Delta\phi/\mathrm{e}$ of the electric potentials of tip and sample is generated. Here, e is the elementary charge and $\Delta\phi = \phi_{\mathrm{s}} - \phi_{\mathrm{t}}$. If in addition a voltage $V_{\mathrm{ts}}$ is applied between tip and sample, $V=V_{\mathrm{ts}} - V_{\Delta\phi}$. Therefore $\Delta\phi$ can be determined by measuring $F_{\mathrm{es}}(V_{\mathrm{ts}})$, which is done by Kelvin probe microscopy.\cite{nonnenmacher} We note that $F_{\mathrm{es}} = 0$ if $V_{\mathrm{ts}} = V_{\Delta\phi}$.

The $F_{\mathrm{es}}(V_{\mathrm{ts}})$ characteristic can be determined by measuring the frequency shift $\Delta f$ of the cantilever resonance frequency as a function of $V_{\mathrm{ts}}$, because for small oscillation amplitudes, $\Delta f(V_{\mathrm{ts}}) \propto \mathrm{d}F_{\mathrm{total}}(V_{\mathrm{ts}})/\mathrm{d}z \propto \mathrm{d}F_{\mathrm{es}}(V_{\mathrm{ts}})/\mathrm{d}z + \mathrm{d}F_{\mathrm{topo}}/\mathrm{d}z$.\cite{albrecht} $F_{\mathrm{total}}(V_{\mathrm{ts}})$ is the total force comprising $F_{\mathrm{es}}(V_{\mathrm{ts}})$ as well as a "topographic force" $F_{\mathrm{topo}}$, which includes forces such as van der Waals and chemical bonding forces between tip and sample. A representative spectrum of $\Delta f(V_{\mathrm{ts}})$ measured on a sample with 4\,unit cells LaAlO$_3$ at 4.7\,K is shown in Fig. 3(a). The $\Delta f(V_{\mathrm{ts}})$ characteristic has a pronounced parabolic dependence (see inset of Fig. 3(a)). $V_{\Delta\phi}$ is given by the voltage of the $\Delta f(V_{\mathrm{ts}})$ parabola's extremum. The measured sign of $V_{\Delta\phi}$ implies that $\phi_{\mathrm{s}}$ is greater than $\phi_{\mathrm{t}}$, as is the case in the illustration shown in Fig. 3(b).

\begin{figure}
  \includegraphics[width=\columnwidth]{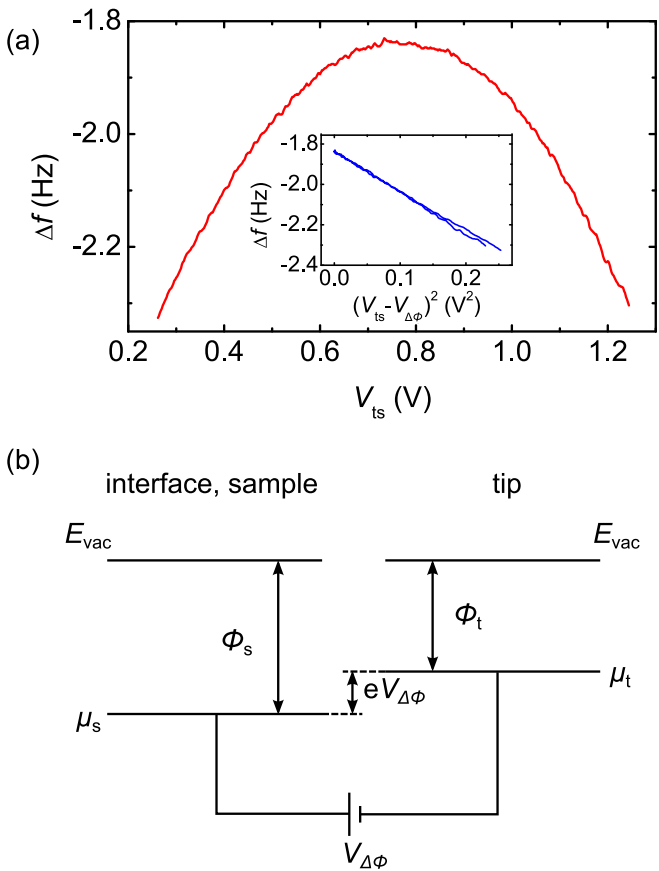}
	\caption{(a) Frequency shift $\Delta f$ of the cantilever oscillation frequency measured as a function of tip--sample voltage $V_{\mathrm{ts}}$. The spectrum was taken at 4.7\,K with a sweep rate of 1\,V/min on a position distant from topographic steps. The linear curve in the inset shows the quadratic dependence of $\Delta f$ on $V_{\mathrm{ts}}-V_{\Delta\phi}$. (b) Schematic band diagram of the interface and the tip. The contact potential difference $V_{\Delta\phi}$ corresponds to the difference of the work functions of interface $\phi_{\mathrm{s}}$ and tip $\phi_{\mathrm{t}}$. The nominal work function of the iridium tip is 5.27\,eV\cite{wilson} and we measure mV changes in $V_{\Delta\phi}$. The electrochemical potentials of sample and tip are labeled $\mu_{\mathrm{s}}$ and $\mu_{\mathrm{t}}$, respectively. In the case shown, the vacuum levels $E_{\mathrm{vac}}$ are balanced by a voltage $V_{\Delta\phi}$ applied between interface and tip.}
  \label{fig:band}
\end{figure}
$\Delta\phi$ is readily obtained from the measurement of $V_{\Delta\phi}$, and therefore also $\phi_{\mathrm{s}}$ if $\phi_{\mathrm{t}}$ is known. To measure the electronic compressibility of the interface, however, $\phi_{\mathrm{s}}$ has to be measured as a function of the interface carrier density $n$. To measure $\Delta\phi(n)$, $F_{\mathrm{es}}$ needs to be measured at constant $z$, while $n$ is varied by applying a gate voltage $V_{\mathrm{G}}$ to the back of the SrTiO$_3$ substrate. To avoid artifacts resulting from the electrostriction of the SrTiO$_3$,\cite{schmidt} for every $V_{\mathrm{G}}$ the tip--sample distance $z$ needs to be adjusted to the same value. This is done by making use of the $\mathrm{d}F_{\mathrm{topo}}/\mathrm{d}z (z)$ dependence: for every $V_{\mathrm{G}}$, the tip biased at $V_{\mathrm{ts}} \approx V_{\Delta\phi}$ is moved in $z$-direction to obtain the same $\mathrm{d}F_{\mathrm{topo}}/\mathrm{d}z$ value.

The results of the $\Delta\phi(V_{\mathrm{G}})$ measurement are shown in Fig. 4(a). Starting at $V_{\mathrm{G}} = 0$, $V_{\Delta\phi}$ is increased if $V_{\mathrm{G}}$ is lowered, i.e. if $n$ is lowered. At $V_{\mathrm{G}} = -70$\,V, the slope of the $V_{\Delta\phi}$ curve changes sign. $V_{\Delta\phi}$ displays a clear minimum at $V_{\mathrm{G}} = -95$\,V. Using a second sample and different tips, we found that this minimum was reproducible. The $V_{\mathrm{G}}$ value of the minimum was found to differ among the samples, which we attribute to differences in the response of the samples on electric fields. The shape and the composition of the tips used in the experiments affected the absolute value of the minimum.

\begin{figure}
  \includegraphics[width=\columnwidth]{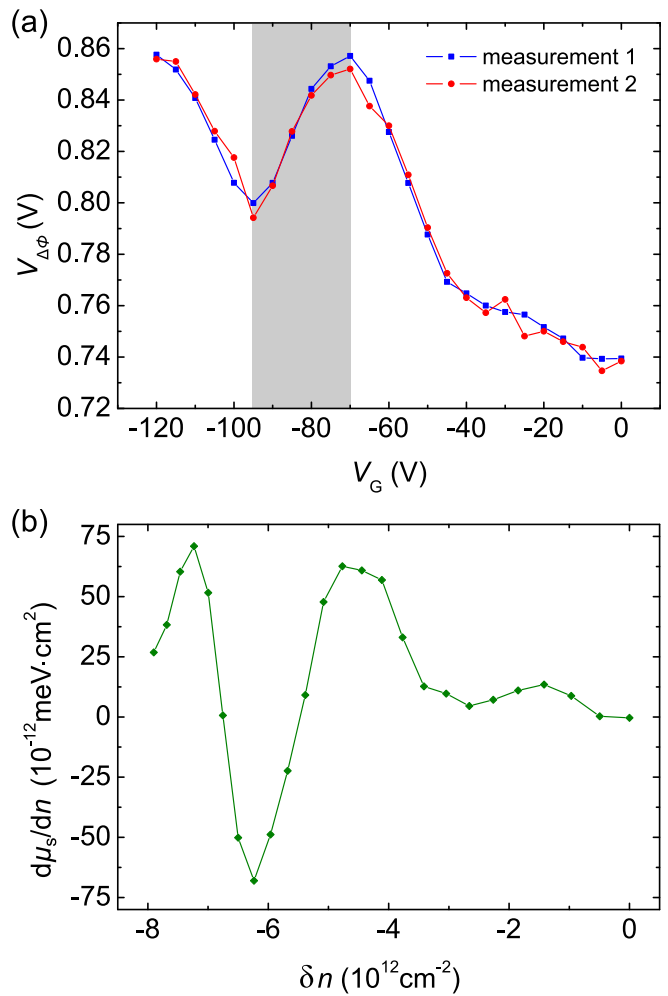}
	\caption{(a) Contact potential difference $V_{\Delta\phi}$ measured as a function of $V_{\mathrm{G}}$. The measurements were performed at 4.7\,K. Spectrum 2 was recorded directly after spectrum 1. $V_{\Delta\phi}$ was obtained through quadratic fits of the $\Delta{}f(V_{\mathrm{ts}})$ spectra. (b) The inverse electronic compressibility $\mathrm{d}\mu_{\mathrm{s}}/\mathrm{d}n$ plotted as function of the change of the carrier density $\delta n$. The value of $\delta n$ was determined by integrating the capacitance between interface and back gate over the gate voltage.}
  \label{fig:meas}
\end{figure}
Because the work function of the tip $\phi_{\mathrm{t}}$ does not depend on $n$, any change of $\Delta\phi(V_{\mathrm{G}})$ is caused by a change of $\phi_{\mathrm{s}}$, resulting from a shift of the interface chemical potential with respect to the vacuum level $E_\mathrm{vac}$. A decrease of $\Delta\phi$, i.e. a decrease of $\phi_{\mathrm{s}}$, corresponds to an increase of the interface chemical potential. Our data show that $V_{\Delta\phi}$ decreases in the voltage range from $-70$\,V to $-95$\,V. Hence in this range $\mathrm{d}\mu_{\mathrm{s}}/\mathrm{d}n <0$, the electronic compressibility is negative.\cite{bello,tanatar}

To obtain the carrier density $n$ of the interface electron system we measured the capacitance $C$ between the interface and the back gate. For this, an AC voltage with a frequency of $f_{\mathrm{mod}} = 77$\,Hz and an amplitude of $V_{\mathrm{mod}} = 10$\,mV$_{\mathrm{rms}}$ was applied to the interface and the resulting current $I$ was measured with a standard lock-in technique. Because leakage currents were found to be negligibly small, the AC current is proportional to $C$. A clear decrease in capacitance with increasing electric field is observed, which is caused by the field dependency of the dielectric constant of SrTiO$_3$.\cite{christen} The change in the carrier density $\delta n$ with applied gate voltage can be calculated by integrating the measured capacitance. The carrier density is the sum of the density at zero gate voltage $n_0$ and $\delta n$: $n=n_0+\delta n$. The inverse compressibility $\mathrm{d}\mu_{\mathrm{s}}/\mathrm{d}n$ can be determined by differentiating $V_{\Delta\phi}$ with respect to $\delta n$: $\mathrm{d}\mu_{\mathrm{s}}/\mathrm{d}n = \mathrm{d}\mu_{\mathrm{s}}/\mathrm{d}(\delta n) = -\mathrm{d}(\mathrm{e}V_{\Delta\phi})/\mathrm{d}(\delta n)$. The result is shown in Fig. 4(b) and exhibits a clear dip of the electronic compressibility at $\delta n = -6.2\times10^{12}$\,/cm$^{2}$. It is of the same order of magnitude as measured by Lu~Li \em et al\em{}.\cite{luli_kap} The absolute value differs slightly from the previously measured data, which is ascribed to the differences in the LaAlO$_3$ film thicknesses.

In summary, we have shown with Kelvin probe microscopy that the conducting LaAlO$_3$-SrTiO$_3$ interface exhibits a negative electronic compressibility at low carrier densities even if the LaAlO$_3$ layer is only four unit cells thick. This effect is consistent with the results of Lu~Li \em et al.\em{},\cite{luli_kap} who found a negative electronic compressibility close to the metal--insulator transition for 10 and 12\,unit cells LaAlO$_3$. The employed technique is independent of the previously used capacitance-measurement method. Moreover, as we deplete the interface solely by applying voltages to the back of the SrTiO$_3$ substrate, the use of a gate on top of the LaAlO$_3$ film is dispensable. It is therefore possible to measure samples with ultrathin LaAlO$_3$ films.

In Ref.~\onlinecite{kopp} it has been suggested that the capacitance of capacitors may be enhanced by optimizing the electrode material. Indeed, Lu~Li \em et al.\em{} found a large enhancement of the capacitance of metal-LaAlO$_3$-SrTiO$_3$ capacitors with LaAlO$_3$ film thicknesses of 10 and 12\,unit cells.\cite{luli_kap} For thinner LaAlO$_3$ films the relative enhancement is proposed to be even larger, as the geometrical capacitance of the device is greater. Owing to tunneling and leakage currents through such thin LaAlO$_3$ films it was not yet possible to build such devices. In view of the results reported above, we suggest to reduce the leakage currents by growing an additional dielectric layer under the top gate on heterostructures with four unit cells LaAlO$_3$. If this additional layer has a much greater dielectric constant than LaAlO$_3$, the geometric capacitance of the device will be reduced by only a small amount. Hence the capacitance enhancement is expected to become even greater than previously observed.

We gratefully acknowledge helpful discussions with T.~Kopp, R.~Jany and K.~Steffen as well as financial support by the Deutsche Forschungsgemeinschaft (TRR~80) and by the European Commission (OxIDes).


\begin{thebibliography}{--}

	\bibitem{ohtomo}
		A.~Ohtomo and H.\,Y.~Hwang, Nature \textbf{427}, 423 (2004).		

	\bibitem{basletic}
		M.~Basletic \em et al.\em, Nature Mat. \textbf{7}, 621 (2008).
	
	\bibitem{caviglia}
		A.\,D.~Caviglia \em et al.\em, Phys. Rev. Lett. \textbf{105}, 236802 (2010).

	\bibitem{nakagawa}
		N.~Nakagawa, H.\,Y.~Hwang, and D.\,A.~Muller, Nat. Mat. \textbf{5}, 204 (2006).
		
	\bibitem{salluzzo}
		M.~Salluzzo \em et al.\em, Phys. Rev. Lett. \textbf{102}, 166804 (2009).
		
	\bibitem{breitschaft}
		M.~Breitschaft \em et al.\em, Phys. Rev. B \textbf{81}, 153414 (2010).
		
	\bibitem{pavlenko}
		G. Berner \em et al.\em, \textit{k}-space mapping of the Fermi surface at the LaAlO$_3$/SrTiO$_3$ interface: charge dichotomy, potential buildup, and O vacancies, unpublished.
		
	\bibitem{shalom}
		M.~Ben~Shalom, A.~Ron, A.~Palevski, and Y.~Dagan, Phys. Rev. Lett. \textbf{105}, 206401 (2010).
	
	\bibitem{dubroka}
		A.~Dubroka \em et al.\em, Phys. Rev. Lett. \textbf{104}, 156807 (2010).

	\bibitem{thiel_field}
		S.~Thiel \em et al.\em, Science \textbf{313}, 1942 (2006).	

	\bibitem{dikin}
		D.\,A.~Dikin \em et al.\em, Phys. Rev. Lett. \textbf{107}, 056802 (2011).
	
	\bibitem{luli_coex}
		Lu~Li, C.~Richter, J.~Mannhart, and R.\,C.~Ashoori, Nature Phys. \textbf{7}, 762 (2011).
		
	\bibitem{bert}
		J.\,A.~Bert \em et al.\em, Nature Phys. \textbf{7}, 767 (2011).

	\bibitem{luli_kap}
		Lu~Li \em et al.\em, Science \textbf{332}, 825 (2011).

	\bibitem{eisenstein_prb}
		J.\,P.~Eisenstein, L.\,N.~Pfeiffer, and K.\,W.~West, Phys. Rev. B \textbf{50}, 1760 (1994).

	\bibitem{kravchenko}
		S.\,V.~Kravchenko, V.\,M.~Pudalov, and S.\,G.~Semenchinsky, Phys. Lett. A \textbf{141}, 71 (1989).	

	\bibitem{ilani}
		S.~Ilani, L.\,A.\,K.~Donev, M.~Kindermann, and P.\,L.~McEuen, Nature Phys. \textbf{2}, 687 (2006).
		
	\bibitem{eisenstein}
		J.\,P.~Eisenstein, L.\,N.~Pfeiffer, and K.\,W.~West, Phys. Rev. Lett. \textbf{68}, 674 (1992).
		
	\bibitem{allison}
		G.~Allison \em et al.\em, Phys. Rev. Lett. \textbf{96}, 216407 (2006).

	\bibitem{kala}
		A.\,S.~Kalabukhov \em et al.\em, Phys. Rev. Lett. \textbf{103}, 146101 (2009).
	
	\bibitem{arras}
		R.~Arras, V.\,G.~Ruiz, W.\,E.~Pickett, and R.~Pentcheva, Phys. Rev. B \textbf{85}, 125404 (2012).
		
	\bibitem{foerg}
		B.~F\"{o}rg, C.~Richter and J.~Mannhart, Appl. Phys. Lett. \textbf{100}, 053506 (2012).
		
	\bibitem{kawasaki}
		M.~Kawasaki \em et al.\em, Science \textbf{266}, 1540 (1994).
		
	\bibitem{koster}
		G.~Koster \em et al.\em, Appl. Phys. Lett. \textbf{73}, 2920 (1998).

	\bibitem{giessibl}
		F.\,J.~Giessibl, Appl. Phys. Lett. \textbf{73}, 3956 (1998).
	
	\bibitem{hembacher}
		S.~Hembacher, F.\,J.~Giessibl, and J.~Mannhart, Science \textbf{305}, 380 (2004).
		
	\bibitem{nonnenmacher}
		M.~Nonnenmacher, M.\,P.~O'Boyle, and H.\,K.~Wickramasinghe, Appl. Phys. Lett. \textbf{58}, 2921 (1991).
		
	\bibitem{albrecht}
		T.\,R.~Albrecht, P.~Gr\"{u}tter, D.~Horne, and D.~Rugar, J. Appl. Phys. \textbf{69}, 668 (1991).	
			
	\bibitem{wilson}
		R.\,G.~Wilson, J. Appl. Phys. \textbf{37}, 3170 (1966).
		
	\bibitem{schmidt}
		G.~Schmidt and E.~Hegenbarth, Phys. Stat. Sol. \textbf{3}, 329 (1963).

	\bibitem{bello}
		M.\,S.~Bello, E.\,I.~Levin, B.\,I.~Shklovskii, and A.\,L.~\'{E}fros, Zh. Eksp. Teor. Fiz. \textbf{80}, 1596 (1981) [Sov. Phys. JETP \textbf{53}, 822 (1981)].
	
	\bibitem{tanatar}
		B.~Tanatar and D.\,M.~Ceperley, Phys. Rev. B \textbf{39}, 5005 (1989).

	\bibitem{christen}
		H.-M.~Christen, J.~Mannhart, E.\,J.~Williams, and Ch.~Gerber, Phys. Rev. B \textbf{49}, 12095 (1994).
		
	\bibitem{kopp}
		T.~Kopp and J.~Mannhart, J. Appl. Phys. \textbf{106}, 064504 (2009).

\end{thebibliography}
\end{document}